\documentclass[journal]{vgtc}                

\graphicspath{{figures/}{pictures/}{images/}{./}} 

\usepackage{times}                     

\usepackage{tabu}                      
\usepackage{booktabs}                  
\usepackage{lipsum}                    
\usepackage{mwe}                       

\usepackage{mathptmx}                  

\usepackage{utfsym}
\usepackage{multirow}
\usepackage{pifont}
\usepackage{enumitem}
\usepackage{amssymb}
\usepackage[normalem]{ulem}
\usepackage{xcolor}
\usepackage{booktabs}
\usepackage{graphicx}
\usepackage{makecell}
\usepackage{tabularx}
\usepackage{array}
\usepackage{threeparttable}

\newcommand{\myul}[1]{%
  {%
    \renewcommand{\ULthickness}{0.25pt}%
    \setlength{\ULdepth}{2pt}%
    \textit{\uline{#1}}%
  }%
}

\definecolor{mybl}{RGB}{100,100,100}

\newcommand{\subarrow}{%
  \raisebox{0.2ex}{{$\triangleright$}}\hspace{0.3em}%
}

\vgtcinsertpkg

\title{MotionAnimator: Semantic-to-Motion Reasoning for Animating Part-Structured 3D Assets in XR}

\author{Jianan Jiang\\ %
        \scriptsize The Pennsylvania State University\\ %
        \scriptsize jnjiang@psu.edu
\and Bin Li\\ %
     \scriptsize The Pennsylvania State University\\ %
     \scriptsize binli@psu.edu
}

\manuscriptnote{\vspace{-1.3cm}}

\teaser{
  \centering
  \includegraphics[width=\linewidth]{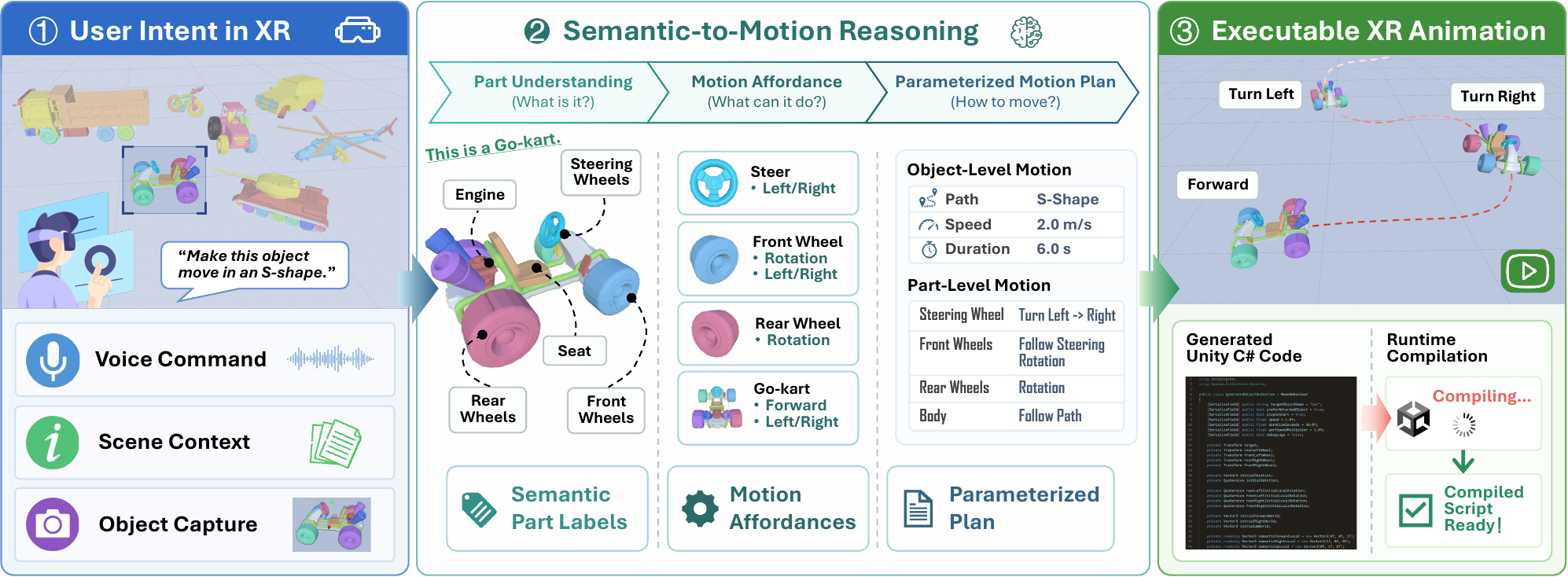}
  \caption{Overview of MotionAnimator. \textcircled{\tiny 1} Given a user’s animation intent, the structured XR scene context, and a target-focused capture of the object, \textcircled{\tiny 2} MotionAnimator performs semantic-to-motion reasoning to identify the object and its functional parts, infer part-specific motion affordances, and construct a parameterized motion plan that coordinates object- and part-level behaviors. \textcircled{\tiny 3} It then generates Unity C\# code, compiles it at runtime, and produces an executable XR animation. In the illustrated example, the go-kart follows an S-shaped path with coordinated steering-wheel motion, front-wheel steering, and wheel rotation.}
  \label{fig:teaser}
}

\abstract{
  3D digital assets are increasingly used in education, simulation, design, and immersive content creation, yet most remain behaviorally inert: they provide geometry and component structure but lack the semantic annotations, motion constraints, and executable behaviors needed to function as dynamic objects in Extended Reality (XR). Animating such assets therefore still requires manual rigging, object-specific templates, or substantial programming expertise. We introduce MotionAnimator, an intent-driven, part-aware XR animation authoring system that transforms semantically unlabeled, part-structured 3D assets into executable animations from natural-language instructions. Rather than generating code directly from text, MotionAnimator externalizes object understanding and motion reasoning into explicit semantic and motion representations. It grounds object identity, semantic parts, and functional directions, constructs a structured plan that coordinates object- and part-level motion, and translates the plan into executable Unity behavior. On a benchmark of 90 prompts across 30 objects, MotionAnimator demonstrated reliable semantic grounding and coordinated animation generation across diverse objects and animation intents. An exploratory user study further provided insights into users' experiences with MotionAnimator and highlighted its potential for part-aware XR animation authoring. These findings demonstrate how explicit semantic and motion representations can bridge high-level user intent and executable behavior, enabling existing part-structured assets to support coordinated XR animations beyond object-level transformations and predefined behaviors.
}


\begin{document}

\maketitle

\section{Introduction}
In recent years, advances in Generative Artificial Intelligence (Generative AI)~\cite{xu2024grm, xiang2025structured, yan2026xpart} have significantly accelerated the development of XR authoring~\cite{numan2024spaceblender, vachha2025dreamcrafter, hou2025echoladder, du2026vibe}. Users can now create realistic virtual objects within immersive environments and interact with them through increasingly natural forms of input~\cite{jiang2026partinteractor, lee2025imaginatear, hu2025thing2reality, chen2025analyzing, zhang2024vrcopilot, dogan2024augmented}. However, relatively little research has explored what happens after these virtual objects are generated, especially how they should move and behave over time. Most existing authoring approaches~\cite{de2024llmr, chen2025llmer, srinidhi2026genassist} for 3D object animation still support only simple functions, such as moving an object from one point to another, rotating it around a target, or triggering a predefined animation. These forms of animation support basic demonstrations but remain insufficient for richer, coordinated object–part behaviors. More importantly, for complex 3D objects, producing functionally plausible animations further requires motion that reflects the object's structure, kinematics, and real-world behavior.

In contrast, animation has already shown substantial value in 2D scenarios~\cite{liu2025logomotion, li2026sketchdynamics}, especially in areas such as education~\cite{chulpongsatorn2023augmented, gunturu2024augmented}, explanation~\cite{kazi2014draco, kazi2016motion}, and visual communication~\cite{subramonyam2018taketoons, rosenberg2024drawtalking}. Animation can help users understand processes, relationships, and dynamic changes that are difficult to convey through static content alone. More broadly, animation is not merely a visual effect, but also a medium for communicating an object's function, state, and possible interactions. In 3D environments, some prior work has explored animation for rigged avatars~\cite{zhou2024timetunnel, yuan2026dancingbox}, virtual characters~\cite{ma2022layered}, and embodied motion authoring~\cite{vogel2018animationvr, arora2019magicalhands}. However, these methods are difficult to generalize across diverse 3D objects because they often rely on predefined skeletons, motion datasets, or category-specific assumptions, while constraining user input and limiting their ability to support open-ended natural language instructions.

For example, a user may select a part-structured object and ask the system to generate a plausible motion behavior, such as “\textit{make the fan turn on},” “\textit{make the drone fly to the table},” or “\textit{animate this object as if it is working}.” Completing such requests is challenging because the system must first interpret the user's high-level intent and then translate that intention into a concrete animation for a specific object. During this process, the system needs to understand what the object is, what parts it contains, what functions it can perform, which parts are movable, and how each part should move in relation to the whole object. These requirements make animating diverse 3D objects much more difficult than applying simple object transformations or triggering predefined behaviors.

The recent development of Large Language Models (LLMs)~\cite{achiam2023gpt, singh2025openai} brings new opportunities to this problem. First, LLMs have demonstrated strong capabilities in understanding multimodal information~\cite{zhang2024omg, huang2024vtimellm, chu20253d, chen2025emos}. This makes them promising tools for interpreting user intent~\cite{hou2025echoladder, bovo2025embardiment}, reasoning about object semantics~\cite{li2024situationadapt}, and analyzing complex scene information~\cite{dogan2024augmented}. Second, LLMs have also shown strong generation capabilities. With the rise of “\textit{vibe coding}”~\cite{ge2025survey, jiang2026survey}, it has become increasingly feasible for users to describe desired behaviors in natural language and obtain executable code in different programming languages~\cite{jennings2024s, xia2025scenegenagent, seo2026paper2code}. Based on these two capabilities, it becomes possible to start from a user's natural language request and generate executable Unity C\# code that controls object animation in XR~\cite{de2024llmr, chen2025llmer, srinidhi2026genassist}.

Nevertheless, directly generating animation code from natural-language instructions introduces significant challenges. Although LLMs can increasingly produce executable code~\cite{wang2022compilable, zhong2024debug, chen2024teaching}, syntactically valid programs may still result in incorrect or implausible animations~\cite{liu2023your, zhang2024naturalcodebench, fakhoury2024llm}. For example, a system may animate a component that should remain fixed, apply motion along an incorrect functional axis, misinterpret the object's structure, or fail to coordinate part-level behavior with object-level motion. \textit{\myul{Therefore}, the central \textbf{challenge} is not merely to generate executable code from language instructions, but to translate high-level animation intent into a semantically grounded and coordinated motion specification for the selected object before realizing it as runtime behavior.}

In this paper, we present MotionAnimator~\footnote{See our \href{https://horizonjohn.github.io/MotionAnimator/}{\textbf{P{\small{ROJECT}} P{\small{AGE}}}} for source code and implementation details.}, an intent-driven, part-aware XR animation authoring system for semantically unlabeled, part-structured 3D assets. Its key idea is to resolve animation intent into explicit asset semantics and coordinated motion specifications before generating runtime code. MotionAnimator decomposes this process into semantic understanding, motion planning, and code generation for runtime execution. Given a natural-language instruction and a selected object, the system first captures visual, spatial, and hierarchical context to infer the object's identity, semantic direction frame, and part labels. It then constructs a structured motion plan that grounds directions and motion axes in the selected asset, distinguishes object-level motion from part-relative motion, and specifies their temporal and coordination relationships. Finally, the plan is translated into constrained Unity C\# code, compiled at runtime, and attached to the selected object. By preserving semantic and motion decisions as explicit intermediate representations, this decomposition supports coordinated object--part animation while making the generation process inspectable and independently evaluable. More broadly, MotionAnimator positions large language models not merely as code generators, but as semantic mediators between high-level user intent and executable XR animation behavior. In summary, our contributions are:

\vspace{-2mm}
\begin{itemize}[leftmargin=2.2em, itemsep=-0.5mm]
    \item We formulate \textit{Semantic Motion Grounding} for existing 3D assets: the problem of translating high-level natural-language intent into asset-specific, coordinated, and executable animation for static, semantically unlabeled, part-structured objects.
    
    \item We introduce \textit{Paired Asset-Grounded Intermediate Representations}: an object-centric semantic abstraction that encodes the object's identity, functional directions, and part labels; and a structured motion specification that defines object- and part-level motions and their coordination dependencies.
    
    \item We present MotionAnimator, a semantic-to-motion-to-execution XR pipeline that combines multimodal asset understanding, asset-specific motion parameterization, and constrained runtime code generation to animate existing 3D assets without manual rigging or predefined templates.
    
    \item We evaluate MotionAnimator on 90 object–prompt cases across 30 objects and in an exploratory 9-participant study, demonstrating the value of explicit semantic and motion reasoning and characterizing users' authoring experiences.
\end{itemize}
\vspace{-3mm}

\begin{figure*}[ht]
    \centering
    \includegraphics[width=1.0\linewidth]{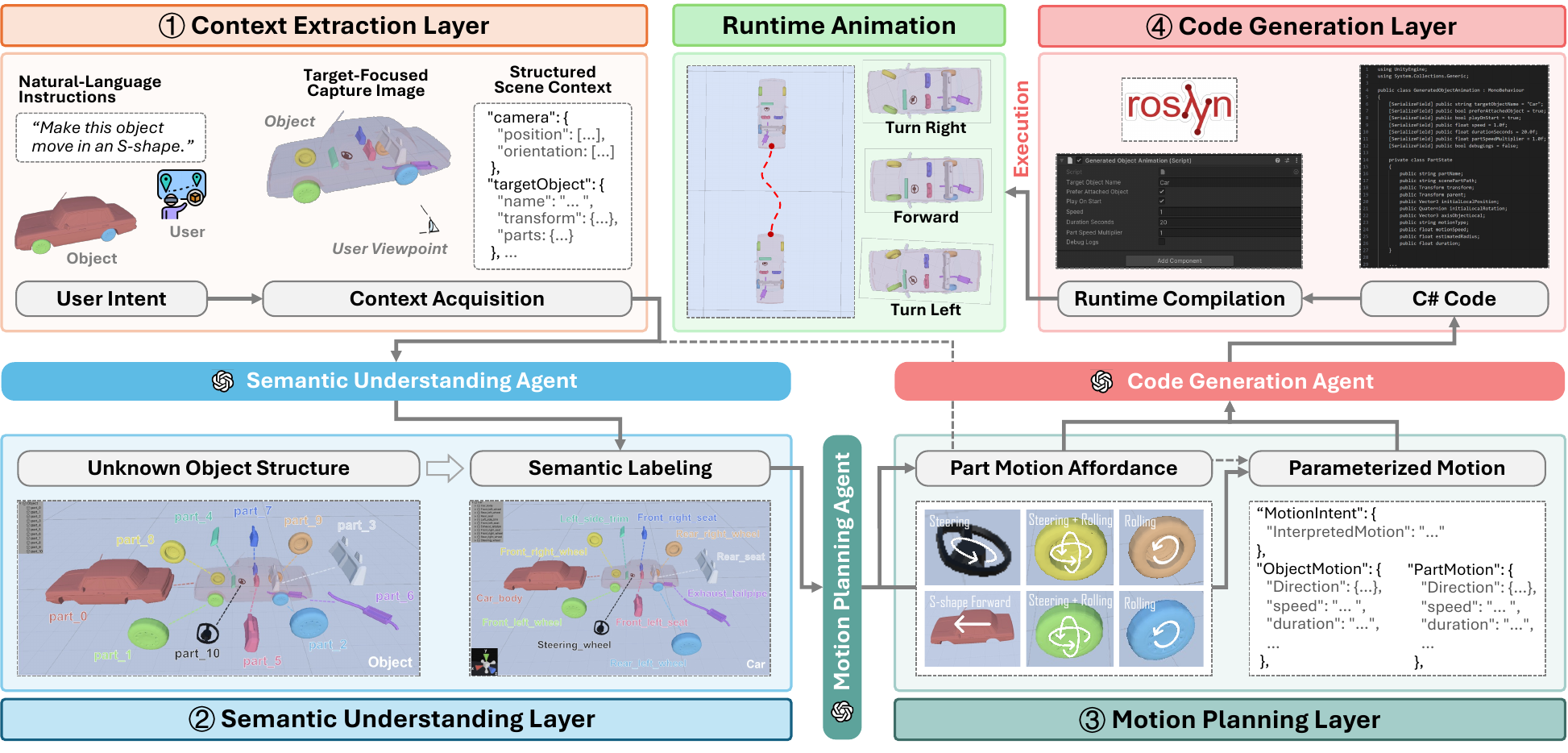}
    \vspace{-6mm}
    \caption{MotionAnimator Architecture. Given a natural-language instruction and a part-structured 3D asset, the system acquires visual and structural context, constructs an object-centric semantic representation, derives a geometry-grounded motion plan that coordinates object- and part-level behavior, and synthesizes executable Unity C\# code for runtime animation.}
    \vspace{-2mm}
    \label{fig:system_overview}
\end{figure*}


\section{Related Work}
\subsection{Generative XR and Animation Authoring}
Generative Artificial Intelligence (Generative AI)~\cite{xu2024grm, xiang2025structured, yan2026xpart} and Large Language Models (LLMs)~\cite{achiam2023gpt, team2023gemini, touvron2023llama, singh2025openai} have substantially lowered the barrier to XR authoring by allowing users to describe desired scenes, objects, layouts, and interactions through natural-language or multimodal input. Prior systems use LLMs to create interactive objects from user prompts~\cite{de2024llmr, giunchi2024dreamcodevr, chen2025llmer, srinidhi2026genassist}, support mixed-initiative spatial layout and progressive scene construction~\cite{zhang2024vrcopilot, hou2025echoladder}, and generate virtual assets grounded in physical or collaborative contexts~\cite{lee2025imaginatear, hu2025thing2reality, numan2024spaceblender}. Other work uses object understanding for context-aware interaction~\cite{dogan2024augmented}, supports immersive editing of complex generated 3D content~\cite{vachha2025dreamcrafter, shen2025gaussianshopvr}, and extends whole-object authoring toward fine-grained interaction with semantic object parts~\cite{jiang2026partinteractor, lee2026hoicraft}. \textit{These systems reflect a progression from scene- and object-level generation toward increasingly interactive, editable, and part-aware XR content.}

Animation authoring research has similarly explored how users can communicate motion intent without specifying every low-level parameter. In 2D, prior systems support coordinated motion across visual elements~\cite{kazi2014draco}, higher-level control of stylized animation~\cite{kazi2016motion}, and workflows connecting performance with animation editing~\cite{subramonyam2018taketoons}. Recent work uses sketches, speech, visual examples, and semantic cues to author interactive animations~\cite{gunturu2024augmented, rosenberg2024drawtalking}, synthesize visually grounded animation code~\cite{liu2025logomotion}, and infer animation intent from free-form input~\cite{li2026sketchdynamics}. In 3D and XR, tracked controllers~\cite{vogel2018animationvr} and mid-air gestures~\cite{arora2019magicalhands} enable direct spatial motion authoring, while spatial-temporal interfaces support organizing and editing complex animations~\cite{ma2022layered, zhou2024timetunnel}. Physical proxies and lightweight motion capture further support motion demonstration and transfer~\cite{li2024anicraft, yuan2026dancingbox}, and AI-assisted interfaces enable refinement of generated avatar motion~\cite{oh2025mowa}. More recently, LLM-based XR systems let users specify object behaviors through natural language by directly generating runtime code~\cite{de2024llmr, giunchi2024dreamcodevr} or translating instructions into structured commands and predefined functions~\cite{chen2025llmer, srinidhi2026genassist}.

\subsection{LLM-based Program Generation}
Large Language Models (LLMs) and multimodal LLMs~\cite{achiam2023gpt, singh2025openai} have expanded beyond text reasoning to multimodal understanding of code~\cite{nam2024using}, images~\cite{zhang2024omg}, videos~\cite{huang2024vtimellm}, and 3D-related representations~\cite{chu20253d}. This makes them increasingly relevant to interactive authoring systems that require semantic interpretation rather than direct manipulation alone. Prior XR and interactive systems use LLMs for progressive scene design~\cite{hou2025echoladder}, embodied assistance~\cite{bovo2025embardiment}, context-aware interface adaptation~\cite{li2024situationadapt}, object-centered XR interaction~\cite{dogan2024augmented}, and scene-wide behavior generation from object semantics and physical context~\cite{jingu2026scene2hap}. \textit{These works demonstrate the potential of LLMs to connect high-level user intent with situated spatial, semantic, and contextual information. }

LLMs have also reshaped executable program generation. Early code-generation models~\cite{chen2021evaluating, li2022competition, nijkamp2023codegen} demonstrated the feasibility of synthesizing code from natural-language specifications, while more recent ``\textit{vibe coding}'' workflows let users describe high-level goals as LLMs generate, revise, and repair implementations~\cite{ge2025survey, jiang2026survey}. This paradigm has influenced interactive and XR authoring through runtime behavior generation~\cite{jennings2024s, de2024llmr, giunchi2024dreamcodevr}, coding-agent-based scene generation~\cite{xia2025scenegenagent, hu2024scenecraft}, JSON-based XR function invocation~\cite{chen2025llmer, patil2024gorilla}, prompt-driven XR program generation~\cite{srinidhi2026genassist}, and WebXR prototyping~\cite{du2026vibe}. However, executable generation also inherits the reliability challenges of LLM-generated code. Prior work has explored compiler feedback~\cite{wang2022compilable}, execution tracing~\cite{zhong2024debug}, and self-debugging with execution results~\cite{chen2024teaching}, while complementary studies emphasize stronger functional evaluation~\cite{liu2023your}, realistic user-query benchmarks~\cite{zhang2024naturalcodebench}, and interactive intent clarification~\cite{fakhoury2024llm}.

\textbf{Taken together}, existing research has advanced both high-level XR authoring and natural-language program generation, but an important gap remains between expressing animation intent and realizing it as asset-specific motion. XR and animation-authoring systems largely focus on scene- or whole-object-level creation, direct behavior generation, or predefined motion structures, while LLM-based program-generation approaches primarily address how high-level instructions can be translated into executable code. For part-structured 3D assets, however, executable behavior additionally requires reasoning about: \textit{Which components should move or remain fixed; How their motions should be grounded in the asset's structure and functional directions; and How part-level behaviors should coordinate with whole-object motion}. MotionAnimator addresses this gap by explicitly grounding asset semantics and constructing coordinated motion specifications before generating runtime behavior.

\section{MotionAnimator}


\subsection{Motivating Challenges}

Animating semantically unlabeled, part-structured assets from natural-language intent raises three key design challenges:

\vspace{-1mm}
\begin{itemize}[leftmargin=8mm, itemsep=-0.5mm]
    \item[\textbf{C1}:] \textit{How can animation-relevant semantics be inferred from a semantically unlabeled, part-structured 3D asset?}

    \item[\textbf{C2}:] \textit{How can high-level functional intent be translated into plausible and coordinated object- and part-level motion?}

    \item[\textbf{C3}:] \textit{How can an asset-grounded motion specification be realized as runtime behavior without reinterpreting or losing its upstream semantic and coordination decisions?}
\end{itemize}
\vspace{-1mm}

These challenges motivate a semantic-to-motion-to-execution pipeline that systematically decomposes the animation process into grounded semantic asset understanding, explicit coordinated motion planning, and executable code generation for runtime behavior.


\subsection{System Overview}
In this work, we introduce MotionAnimator, an intent-driven XR animation authoring system that transforms semantically unlabeled, part-structured 3D assets into coordinated runtime animations from natural-language instructions. Such assets may come from online repositories, shared libraries, or existing design workflows, where they contain separable components but lack semantic labels, animation rigs, or predefined behaviors. 

As shown in Figure~\ref{fig:system_overview}, MotionAnimator comprises a four-stage pipeline: (i)~\myul{Context Extraction Layer} acquires the instruction together with visual and structural evidence about the selected object and XR scene. (ii)~\myul{Semantic Understanding Layer} infers the object identity, semantic direction frame, and part labels. (iii)~\myul{Motion Planning Layer} determines what should move, how it should move, and how object- and part-level motions should be coordinated. (iv)~\myul{Code Generation Layer} translates the grounded plan into a constrained Unity C\# behavior for runtime compilation and execution.

\subsubsection{Context Extraction Layer}
The workflow begins when the user selects an object in the XR scene and provides an animation instruction, such as ``\textit{make this object move in an S-shape}.'' MotionAnimator transcribes the instruction, resolves the selected object, and captures a target-focused image from the user's current viewpoint. 

Because the multimodal agent cannot directly inspect the selected 3D file or its complete geometry and hierarchy, MotionAnimator represents the authoring context through two complementary inputs: (i) a target-focused image and (ii) a structured scene context. The latter contains the camera pose, selected-object world transform and object-local bounds, and the hierarchy path, transform, dimensions, and relative size of each component. The image provides visual cues about appearance, orientation, and functional shape, while the structured context preserves component relationships and spatial information unavailable from a single view.

\begin{figure}[t]
    \centering
    \includegraphics[width=1.0\linewidth]{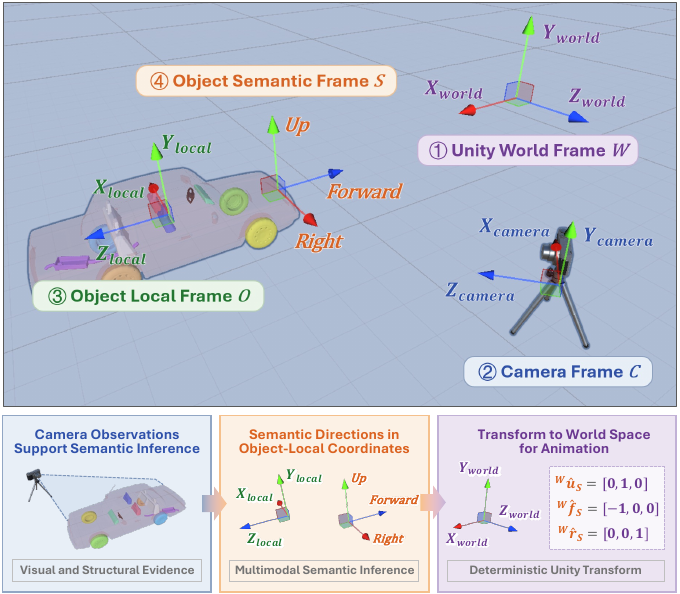}
    \vspace{-6mm}
    \caption{Semantic-Grounded Coordinate Frame. MotionAnimator infers functional directions in the object-local frame and deterministically transforms them into Unity world space, enabling consistent interpretation of directional intent across imported orientations.}
    \vspace{-3mm}
    \label{fig:part_understand}
\end{figure}

\subsubsection{Semantic Understanding Layer}
Using the target-focused image and structured scene context, a \textit{Semantic Understanding Agent} derives an animation-relevant representation of the selected asset. It jointly reasons over visual appearance, component geometry, spatial arrangement, and hierarchy to infer the object identity, establish an object-centric semantic direction frame, and assign semantic labels to components whose original names may be missing, arbitrary, or functionally uninformative. 

A key component of this representation is the semantic direction frame. As shown in Figure~\ref{fig:part_understand}, MotionAnimator distinguishes four coordinate frames: (i) Unity World Frame \(W\), defining the global XR scene; (ii) Camera Frame \(C\), representing the captured viewpoint; (iii) Object Local Frame \(O\), determined by the imported asset transform; and (iv) Object Semantic Frame \(S\), whose axes encode the object's functional forward, right, and up directions. The imported local frame alone is insufficient because its axes may not align with the object's functional orientation.

The agent therefore infers the semantic frame from visual and structural evidence. Semantic forward, right, and up directions, denoted by \(d \in \{f, r, u\}\), are represented as normalized vectors \(^{O} d_{S}\) in the selected-object local frame and deterministically converted into Unity world space. These directions provide a semantic reference basis for trajectories and runtime motion, allowing instructions such as ``forward,'' ``turn left,'' or ``move upward'' to be interpreted according to the object's functional orientation regardless of how the asset was modeled, imported, or rotated. 

The agent also analyzes each component's appearance, relative position, dimensions, repetition, and neighboring relationships to infer its functional role. We formulate the resulting asset-grounded semantic representation as $R_s=\langle o,F_S,P,\phi\rangle$, where $o$ denotes the object identity, $F_S=\langle{}^{O}f_S,{}^{O}r_S,{}^{O}u_S\rangle$ is the semantic direction frame, $P$ its semantic parts, and $\phi:P\rightarrow H$ maps each semantic part to a node in the asset hierarchy $H$. Unity realizes these mappings to transform-preserving wrapper nodes centered on the corresponding part bounds, preserving the original meshes and local transforms while providing stable semantic references for subsequent planning and execution.

\begin{figure}[t]
    \centering
    \includegraphics[width=1.0\linewidth]{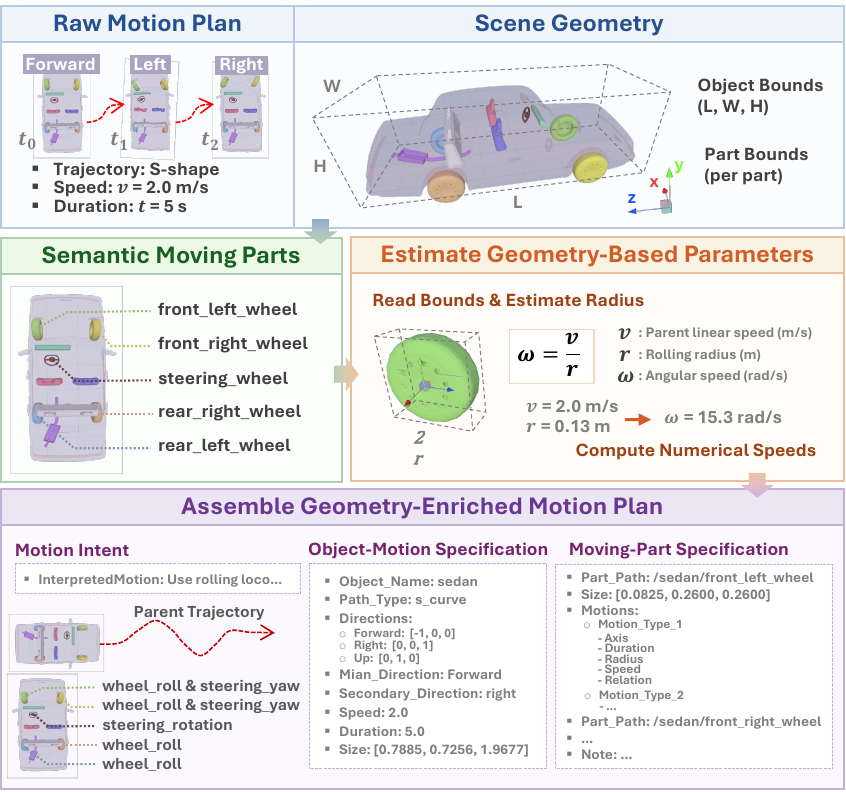}
    \vspace{-6mm}
    \caption{Geometry-Enriched Motion Plan. MotionAnimator uses asset-specific geometry and explicit kinematic relationships to parameterize object- and part-level motion, producing coordinated and functionally plausible animations for subsequent C\# generation.}
    \vspace{-3mm}
    \label{fig:motion_plan}
\end{figure}

\subsubsection{Motion Planning Layer}
Using the original user instruction and semantic hierarchy, a \textit{Motion Planning Agent} converts the requested behavior into a structured motion plan that separates object-level behavior from component-relative motion, identifies independently moving components, and specifies coordination dependencies among their motions. 

We represent the raw motion plan as $R_m=\langle I,M_o,M_p,D\rangle$, comprising the interpreted motion intent $I$, object-level motion $M_o$, part-level motions $M_p$, and their coordination dependencies $D$. Here, $I$ summarizes the intended overall behavior, while $D$ captures qualitative motion relationships and parameter dependencies among object- and part-level motions. The plan further specifies path type, semantic directions, temporal parameters, orientation behavior, independently moving parts, and their motion types and axes. Each movable component may contain one or more concurrent motions.

As illustrated in Figure~\ref{fig:motion_plan}, the raw plan remains primarily semantic and symbolic and therefore lacks some asset-specific information required for execution. MotionAnimator further enriches the plan with geometry extracted from the Unity scene, resolving semantic moving-part names to concrete hierarchy paths and incorporating object- and part-level bounds to support asset-specific motion parameterization. We express this transformation as $R_m^{*}=E(R_m,G)$, where $G$ denotes the asset geometry and $E$ resolves hierarchy references and derives geometry-dependent quantities such as motion radii and numerical speeds. The resulting execution-ready representation preserves the planned kinematic dependencies and serves as the sole input to code generation.

\subsubsection{Code Generation Layer}
A \textit{Code Generation Agent} realizes the execution-ready representation as runtime behavior, $B=C(R_m^{*})$, by translating the specified motion semantics, parameters, and dependencies into a standalone Unity \texttt{MonoBehaviour}. Importantly, it receives only $R_m^{*}$ rather than the original scene context, preserving upstream semantic and motion decisions while allowing the concrete runtime implementation to be synthesized at this final stage.

The generated behavior realizes object-level and part-relative motion according to the coordinate frames, parameters, and dependencies encoded in the plan. Semantic directions govern world-space trajectory and orientation updates, while component motions are applied through the corresponding semantic nodes, preserving dependencies such as rolling, steering, and concurrent part motion. 

Before execution, MotionAnimator normalizes and validates the generated source, compiles it in memory using Roslyn~\footnote{\url{https://github.com/dotnet/roslyn}}, removes any previously generated animation component, and attaches the compiled \texttt{MonoBehaviour} to the selected object. The user can then preview the resulting animation in the XR scene and provide a revised instruction for further refinement.

\begin{figure}[t]
    \centering
    \includegraphics[width=1.0\linewidth]{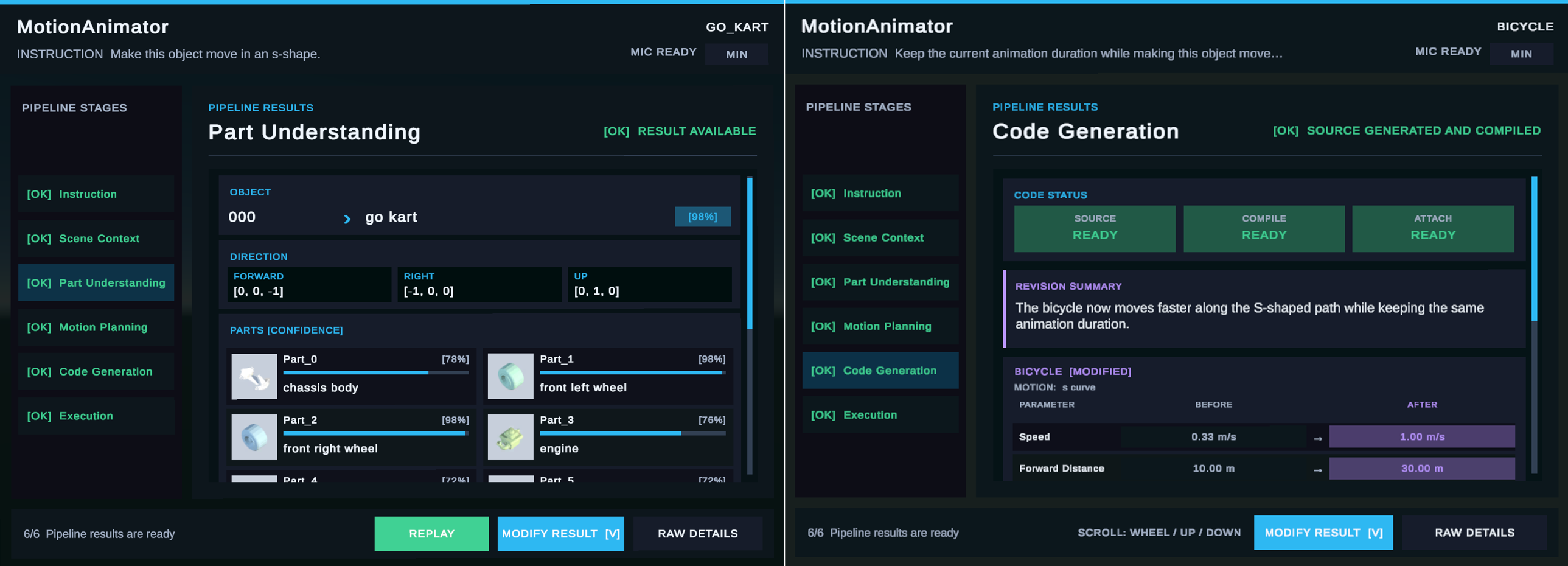}
    \vspace{-6mm}
    \caption{Inspectable and Refinable User Animation Interface.}
    \vspace{-2mm}
    \label{fig:ui}
\end{figure}

\subsection{User Interface}
As shown in Figure~\ref{fig:ui}, MotionAnimator exposes its semantic-to-motion pipeline through a collapsible world-space interface with six stage-aligned views: \textit{User Instruction}, \textit{Scene Context}, \textit{Part Understanding}, \textit{Motion Planning}, \textit{Code Generation}, and \textit{Execution}. Rather than presenting a black-box result, these views allow users to inspect how their instruction is progressively grounded into asset semantics, planned motion, and executable behavior. Each view presents concise, task-relevant information, including the captured scene context, inferred object and part semantics, semantic directions, planned motion types, axes and speeds, and generation and execution status. Additional details can be expanded on demand, keeping the default interface compact while preserving access to information useful for diagnosing semantic or motion mismatches.

Refinement closes this inspection loop. When the user activates \textit{Modify Result} and provides a spoken correction, a \textit{Code Revision Agent} receives the correction together with the original instruction, accepted revision history, geometry-enriched motion plan, and current C\# source. The agent first determines whether the correction preserves the existing motion structure: minor refinements directly revise the code, avoiding a full re-planning pass while retaining the accepted motion structure, whereas structural changes update the motion plan for code revision. This context also supports incremental and referential revisions, such as ``faster than before'' or ``undo the previous change,'' without restating the complete animation intent. The revised code is accompanied by a concise before-and-after summary for verification. Once validated and compiled, MotionAnimator restarts the animation, allowing users to inspect the revision and continue iterating when needed.

\subsection{Implementation Details}
MotionAnimator was implemented in Unity~\footnote{\url{https://unity.com/products/unity-engine}} 2022.3.58f1 using C\# and deployed on a Meta Quest 3~\footnote{\url{https://www.meta.com/quest/quest-3/}}. The Meta XR All-in-One SDK and OpenXR support XR interaction. Speech is recorded in Unity and transcribed through the OpenAI transcription API~\cite{radford2023robust}. For each request, MotionAnimator captures a \(768 \times 538\) target-focused image from the user's viewpoint. During capture, non-target objects are hidden, while the largest component of the target asset is rendered semi-transparently to reveal otherwise occluded internal parts; all objects remain active in the scene and are represented in the structured context. Unity exports the camera pose, target-object transform and bounds, and each component's hierarchy path, dimensions, and relative size as structured JSON. A Python backend communicates with Unity through TCP and invokes the OpenAI API. Initial generation uses three stage-specific agents for semantic understanding, motion planning, and code generation, all using GPT-5.6-Sol~\cite{openai2026gpt56} as the underlying reasoning model; a separate revision agent handles iterative corrections.


\section{Technical Evaluation}
We conducted a technical evaluation to examine whether MotionAnimator can transform high-level natural-language intent into plausible, part-aware, and executable XR animations across diverse semantically unlabeled, part-structured 3D objects.

\subsection{Evaluation Goals}
We evaluate the system from four complementary perspectives:

\vspace{-2mm}
\begin{itemize}[leftmargin=9mm, itemsep=-1mm]
    \item[\textbf{TQ1}:] \textbf{Cross-Object Effectiveness.}
    \textit{Can MotionAnimator generate executable and coordinated animations across diverse instructions and objects that vary in functional category, component structure, and motion affordance?} 
    
    \item[\textbf{TQ2}:] \textbf{Semantic Understanding Accuracy.}
    \textit{How accurately does MotionAnimator identify the selected object, its semantic parts, and its functional direction frame?}

    \item[\textbf{TQ3}:] \textbf{Motion Planning Quality.}
    \textit{How well does MotionAnimator translate user intent into plausible and coordinated object- and part-level motion plans?}

    \item[\textbf{TQ4}:] \textbf{Code and Animation Realization.}
    \textit{How effectively does MotionAnimator translate the generated motion plan into executable code and runtime animation that satisfy the original user instruction?}
\end{itemize}
\vspace{-2mm}

To address these questions, we constructed a benchmark that supports progressive evaluation of the actual intermediate and final outputs produced by the complete MotionAnimator pipeline.

\begin{table}[t]
\centering
\caption{Technical evaluation benchmark, which includes 30 rigid, part-structured 3D assets across five motion-relevant categories, covering whole-object locomotion, steering, continuous rotation, hinge articulation, linear sliding, and coordinated object--part motion.}
\label{tab:object_set}
\begin{threeparttable}
\resizebox{\linewidth}{!}{
\begin{tabular}{lcl}
\toprule
\textbf{Category}
& \textbf{Assets}
& \textbf{Objects} \\
\midrule

Ground Mobility
& 12
& \makecell[l]{
Tractor; Tank; Truck; Bicycle; Motorcycle; Scooter \\
Car$^\dagger$: Hatchback; Sedan; SUV; Pickup; Go-kart; Van}
\\ \midrule

Aerial Mobility
& 6
& \makecell[l]{
Helicopter; Airplane \\
Drone$^\dagger$: Integrated Quadcopter; Guarded Quadcopter; \\
Fuselage Quadcopter; Radial Octocopter}
\\ \midrule

Aquatic Mobility
& 3
& Rowboat; Speedboat; Submarine
\\ \midrule

Rotating Appliances
& 4
& \makecell[l]{
Washing Machine; Record Player \\
Fan$^\dagger$: Table Fan; Ceiling Fan}
\\ \midrule

\makecell[l]{
Articulated \& \\ 
Sliding Objects}
& 5
& \makecell[l]{
Laptop; Drawer Unit \\
Door$^\dagger$: Revolving Door; Double-Hinged Door; Sliding Door}
\\

\midrule
\textbf{Total}
& \textbf{30}
& -- \\
\bottomrule
\end{tabular}
}

\begin{tablenotes}
\small
\item {Note: Fine-grained structural variants within an object family are marked with $^\dagger$.}
\end{tablenotes}

\end{threeparttable}
\end{table}

\subsection{Benchmark Construction}
\subsubsection{Object Set}
As summarized in Table~\ref{tab:object_set}, the benchmark contains 30 part-structured assets spanning five motion-relevant categories. The collection captures diversity at two levels: coarse-grained variation across functionally distinct object categories and fine-grained variation among structurally different assets within the same object family. Among the 30 assets, 12 were sourced from open-source 3D datasets~\cite{yan2026xpart} and 18 from the Unity Asset Store~\footnote{\url{https://assetstore.unity.com/3d}}.

\begin{figure}[t]
    \centering
    \includegraphics[width=1.0\linewidth]{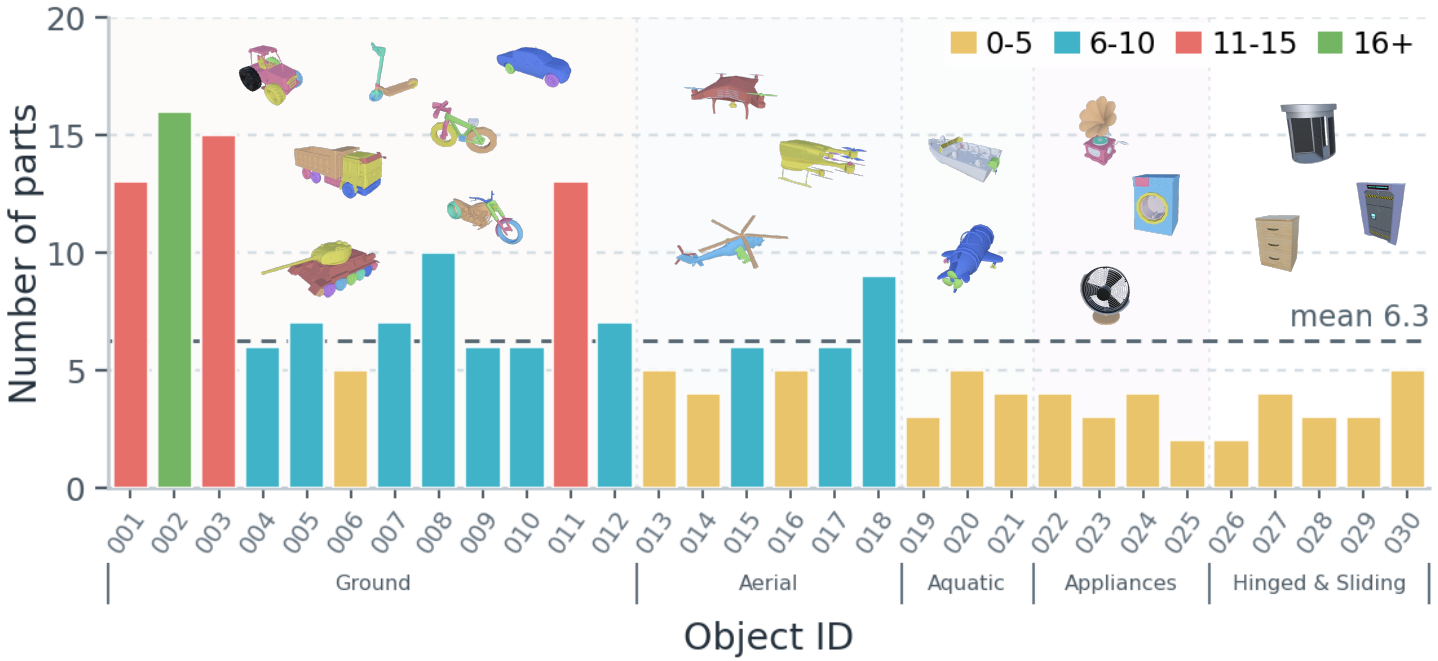}
    \vspace{-6mm}
    \caption{Component counts of the benchmark assets, grouped by motion-relevant category. Bar colors indicate component-count ranges, and thumbnails show representative assets.}
    \vspace{-2mm}
    \label{fig:benchmark}
\end{figure}

As shown in Figure~\ref{fig:benchmark}, each asset contains 2--18 hierarchically separated rigid components (Mean = 6.3) and supports at least one plausible object- or part-level motion affordance. The object set therefore spans both compact structures and more complex assets with repeated or coordinated parts. To prevent semantic leakage, we anonymized all original part names while preserving each asset's original local coordinate frame and mesh hierarchy.

\subsubsection{Prompt and Annotation Design}
For each asset, we authored three prompts at different levels of granularity, yielding 90 asset--prompt cases. The prompts represent three forms of motion specification:
(i) \myul{Simple Motion Prompts} express a direct request with minimal constraints; 
(ii) \myul{Constrained Motion Prompts} add spatial, temporal, trajectory, or action constraints; and 
(iii) \myul{Open-Ended Prompts} use the same generic instruction across all assets---``\textit{Animate this object in a plausible way based on how it normally functions}.''---requiring the system to infer an appropriate behavior from the object's semantics and functional affordances without explicit motion guidance.

Two annotators independently created reference specifications for the benchmark. For each asset, (i) \myul{Semantic Specification} defined its object identity, semantic part labels, and semantic direction frame. For each asset--prompt case, (ii) \myul{Motion Specification} defined the expected moving parts and object- and part-level motions, including motion type and axis. Disagreements were resolved through discussion to establish a consolidated reference specification capturing the expected core behavior for evaluation.

\subsubsection{Evaluation Protocol}
Before evaluation, we fixed all agent prompts, output schemas, and model inference settings. Each asset--prompt case was processed without manual correction or selective regeneration. Only transient API or communication failures were retried with identical inputs; malformed outputs, compilation failures, and incorrect animations were retained as system outcomes.

For each case, we recorded the semantic representation, motion plan, generated Unity C\# source, and runtime animation. Reference-based metrics were used for properties with consistently annotatable targets. Object identities and part labels were evaluated by semantic equivalence rather than exact string matching, allowing synonymous or linguistically equivalent descriptions. In contrast, direction frames and motion axes were evaluated against the annotated object-local axes and signs. Higher-level qualities that admit multiple plausible realizations, including overall motion interpretation and object--part coordination, were assessed through independent human and LLM judgments.

\subsection{Stage-Wise Evaluation}
To examine \textbf{TQ2} and \textbf{TQ3}, we evaluated the intermediate semantic representation and motion plan produced during each execution.

\textbf{Semantic Understanding.}
We report \underline{OIA} (\textit{Object Identity Accuracy}) to measure whether the predicted semantic object identity matches the annotation; \underline{PLA} (\textit{Part Label Accuracy}) to measure the correctness of semantic labels assigned to the annotated parts; and \underline{DFA} (\textit{Direction Frame Accuracy}) to measure whether the predicted semantic forward, right, and up directions match the annotated object-local axes and signs.

\textbf{Motion Planning.}
We report \underline{MPF1} (\textit{Movable Part F1}) for selecting parts that should participate in the requested behavior; \underline{MAA} (\textit{Motion Axis Accuracy}) for grounding their motion axes or directions; and \underline{IMA} (\textit{Intent-to-Motion Alignment}), a 7-point rating by five independent human raters that assesses the complete motion plan with respect to the original instruction, considering object-level behavior, part-level motion, and their coordination.

\begin{table}[t]
\centering
\caption{Stage-wise Semantic-Understanding (SUA) and Motion-Planning (MPA) performance across benchmark categories.}
\label{tab:agenteval}
\begin{threeparttable}
\resizebox{\linewidth}{!}{
\begin{tabular}{clcccccc}
\toprule
& \textbf{Metric}
& \makecell{\textbf{Ground}\\\textbf{Mobility}}
& \makecell{\textbf{Aerial}\\\textbf{Mobility}}
& \makecell{\textbf{Aquatic}\\\textbf{Mobility}}
& \makecell{\textbf{Rotating}\\\textbf{Appliances}}
& \makecell{\textbf{A \& S}\\\textbf{Objects}}
& \textbf{Overall} \\
\midrule

\multirow{4}{*}{\rotatebox{90}{\textbf{SUA}}}
& OIA
& 94.44\% & 100.00\% & 100.00\% & 100.00\% & 100.00\% & 97.78\% \\

& PLA
& 95.27\% & 95.37\% & 100.00\% & 97.92\% & 100.00\% & 96.90\% \\

& DFA
& 100.00\% & 96.30\% & 100.00\% & 100.00\% & 100.00\% & 99.26\% \\

\cmidrule(lr){2-8}
& \textbf{Avg.}
& 96.57\% & 97.22\% & 100.00\% & 99.31\% & 100.00\% & \textbf{97.98\%} \\

\midrule

\multirow{4}{*}{\rotatebox{90}{\textbf{MPA}}}
& MPF1
& 99.38\% & 98.15\% & 97.78\% & 100.00\% & 100.00\% & \textbf{99.16\%} \\

& MAA
& 96.27\% & 75.00\% & 92.59\% & 100.00\% & 70.00\% & \textbf{87.77\%} \\


\cmidrule(lr){2-8}
& IMA
& 6.07 $\pm$ 0.91 
& 5.30 $\pm$ 1.40 
& 6.09 $\pm$ 0.82 
& 6.22 $\pm$ 0.80 
& 5.45 $\pm$ 1.44 
& \textbf{5.83 $\pm$ 1.16} \\

\bottomrule
\end{tabular}
}

\begin{tablenotes}
\small
\item {Note: IMA ratings range from 1 = Does-Not-Match-At-All to 7 = Fully-Matches.}
\end{tablenotes}

\end{threeparttable}
\vspace{-2mm}
\end{table}

\subsubsection{Results}
As shown in Table~\ref{tab:agenteval}, the \textit{Semantic Understanding Agent} (SUA) achieved consistently high performance across object identity (97.78\%), part labeling (96.90\%), and semantic direction-frame grounding (99.26\%). The near-perfect direction-frame accuracy indicates that MotionAnimator can robustly establish functional object directions despite variations in asset structure and imported orientation, while the remaining part-labeling errors primarily involved visually or structurally similar components.

A clearer gap emerged in the \textit{Motion Planning Agent} (MPA), which achieved 99.16\% MPF1 but 87.77\% MAA, indicating that identifying “\emph{which parts should move}” was substantially easier than grounding “\emph{how they should move}”. Motion-axis errors were most pronounced for Aerial Mobility (75.00\%) and Articulated \& Sliding Objects (70.00\%), whereas Rotating Appliances achieved 100.00\% on both metrics. Human judgments of the complete motion plans showed a similar pattern, with an overall \textit{Intent-to-Motion Alignment} (IMA) of $5.83 \pm 1.16$ and the lowest rating for Aerial Mobility ($5.30 \pm 1.40$). Inter-rater reliability was good for IMA, with five-rater average-measure absolute agreement of $ICC(2,5)=.794$. Overall, these results suggest that MotionAnimator reliably identifies object semantics and behaviorally relevant parts, while precise motion-axis and articulation grounding remains the primary challenge for geometrically complex behaviors.

\subsection{Code and Runtime Realization Evaluation}
To evaluate \textbf{TQ4}, we evaluated both the generated Unity C\# code and the resulting runtime animations from two complementary perspectives.
(i) \textbf{Human Animation Evaluation}: 
We recruited five independent human raters, including two VR developers, two instructors, and one student, to evaluate the generated animations. For each test case, raters viewed the animation from three complementary viewpoints and assessed four dimensions on a 7-point scale: 
\underline{OLM} (\textit{Object-Level Motion}) measures the correctness of the object's overall motion; 
\underline{PLM} (\textit{Part-Level Motion}) measures whether the required parts move appropriately; 
\underline{MD} (\textit{Motion Direction}) measures whether the motion directions or axes are correct; and 
\underline{OPC} (\textit{Object--Part Coordination}) measures whether object- and part-level motions are properly coordinated. 
We first averaged the five ratings for each object--prompt case and then report the mean and standard deviation across cases.
(ii) \textbf{LLM Code Evaluation}: 
We further employed three independent LLM judges as complementary, GPT-5.6-Sol~\cite{openai2026gpt56}, Gemini-3.6-Flash~\cite{google2026gemini36flash}, and Claude-Opus-5~\cite{anthropic2026opus5}, to evaluate the behavioral fidelity of the generated Unity code. Each judge was provided with the original user instruction, target-object image, generated motion plan, and generated C\# code. The original instruction served as the primary evaluation criterion, while the motion plan was used only as supporting context.

\begin{table}[t]
\centering
\caption{Runtime animation quality (Human) and code-level behavioral fidelity (LLM) across benchmark categories.}
\label{tab:animationeval}
\begin{threeparttable}
\resizebox{\linewidth}{!}{
\begin{tabular}{clcccccc}
\toprule
& \textbf{Metric}
& \makecell{\textbf{Ground}\\\textbf{Mobility}}
& \makecell{\textbf{Aerial}\\\textbf{Mobility}}
& \makecell{\textbf{Aquatic}\\\textbf{Mobility}}
& \makecell{\textbf{Rotating}\\\textbf{Appliances}}
& \makecell{\textbf{A \& S}\\\textbf{Objects}}
& \textbf{Overall} \\
\midrule

\multirow{5}{*}{\rotatebox{90}{\textbf{Human}}}
& OLM
& 5.82 $\pm$ 1.22 
& 4.72 $\pm$ 1.81 
& 5.51 $\pm$ 1.63 
& 6.13 $\pm$ 1.11 
& 5.15 $\pm$ 1.86 
& 5.50 $\pm$ 1.57 \\

& PLM
& 5.71 $\pm$ 1.34 
& 4.86 $\pm$ 1.61 
& 5.60 $\pm$ 1.51 
& 6.12 $\pm$ 1.12 
& 4.93 $\pm$ 2.07 
& 5.45 $\pm$ 1.59 \\

& MD
& 5.95 $\pm$ 1.23 
& 5.01 $\pm$ 1.83 
& 5.84 $\pm$ 1.43 
& 6.28 $\pm$ 0.92 
& 5.24 $\pm$ 1.80 
& 5.68 $\pm$ 1.52 \\

& OPC
& 5.49 $\pm$ 1.34 
& 4.79 $\pm$ 1.54 
& 5.31 $\pm$ 1.62 
& 6.18 $\pm$ 0.95 
& 4.91 $\pm$ 1.97 
& 5.33 $\pm$ 1.55 \\

\cmidrule(lr){2-8}
& \textbf{Avg.}
& 5.74 $\pm$ 1.21 
& 4.84 $\pm$ 1.58 
& 5.57 $\pm$ 1.48 
& 6.18 $\pm$ 0.97 
& 5.06 $\pm$ 1.88 
& \textbf{5.49 $\pm$ 1.49} \\

\midrule

\multirow{4}{*}{\rotatebox{90}{\textbf{LLM}}}
& GPT
& 6.33 $\pm$ 0.89
& 6.39 $\pm$ 0.98
& 6.89 $\pm$ 0.33
& 6.25 $\pm$ 0.97
& 6.27 $\pm$ 1.03
& 6.38 $\pm$ 0.91 \\

& Gemini
& 6.97 $\pm$ 0.17
& 6.78 $\pm$ 0.65
& 7.00 $\pm$ 0.00
& 7.00 $\pm$ 0.00
& 7.00 $\pm$ 0.00
& 6.94 $\pm$ 0.31 \\

& Claude
& 6.08 $\pm$ 0.73
& 5.94 $\pm$ 0.64
& 6.11 $\pm$ 0.33
& 6.00 $\pm$ 0.74
& 6.07 $\pm$ 0.70
& 6.04 $\pm$ 0.67 \\

\cmidrule(lr){2-8}
& \textbf{Avg.}
& 6.46 $\pm$ 0.49
& 6.37 $\pm$ 0.66
& 6.67 $\pm$ 0.17
& 6.42 $\pm$ 0.41
& 6.44 $\pm$ 0.54
& \textbf{6.46 $\pm$ 0.50} \\

\bottomrule
\end{tabular}
}

\begin{tablenotes}
\small
\item {Note: Ratings range from 1 = Fundamentally-Incorrect to 7 = Fully-Appropriate.\newline \phantom{Note: }Five-rater ICC(2,5): OLM=.809; PLM=.752; MD=.804; and OPC=.790.}
\end{tablenotes}

\end{threeparttable}
\vspace{-2mm}
\end{table}

\begin{table*}[t]
\centering
\caption{End-to-end animation generation results across baselines, ablations, and the full MotionAnimator pipeline.}
\label{tab:execution}
\begin{threeparttable}

\resizebox{\linewidth}{!}{
\begin{tabular}{l@{\hspace{2.5pt}}c@{\hspace{3pt}}*{14}{c}}
\toprule

\multirow{2}{*}{\textbf{Method}}
& \multirow{2}{*}{\makecell[c]{\textbf{Compilation}\\\textbf{Success (\%)}}}
& \multicolumn{4}{c}{\textbf{Time (s)}}
& \multicolumn{3}{c}{\textbf{Token}}
& \multicolumn{7}{c}{\textbf{Animation Quality}} \\

\cmidrule(lr){3-6}
\cmidrule(lr){7-9}
\cmidrule(lr){10-16}

&
& SUA & MPA & CGA & End-to-End
& Input & Output & All
& OLM & PLM & OPC & Avg. & 95\% CI & $p_{\mathrm{Holm}}$ & $r$ \\

\midrule

Direct2C
& 100.00
& -- & -- & 30.42 & 30.66
& 2,220 & 1,715 & 3,934
& 4.02 $\pm$ 1.42 
& 3.53 $\pm$ 1.90 
& 3.53 $\pm$ 1.75 
& 3.69 $\pm$ 1.62 
& $[3.09, 4.30]$ 
& $.0024$ & $0.63$ \\

All-in-One
& 100.00
& -- & -- & 47.84 & 48.66
& 8,103 & 3,079 & 11,181
& 4.22 $\pm$ 1.51 
& 3.70 $\pm$ 1.64 
& 3.72 $\pm$ 1.65 
& 3.88 $\pm$ 1.52 
& $[3.31, 4.45]$ 
& $.0021$ & $0.64$ \\

\midrule

w/o SUA
& 100.00
& -- & 20.59 & 50.66 & 72.38
& 4,805 & 4,579 & 9,384
& 4.47 $\pm$ 1.35 
& 3.97 $\pm$ 1.71 
& 4.00 $\pm$ 1.58 
& 4.14 $\pm$ 1.47 
& $[3.59, 4.69]$ 
& $.0042$ & $0.58$ \\

w/o MPA
& 100.00
& 19.93 & -- & 42.20 & 62.67
& 5,704 & 3,450 & 9,154
& 2.97 $\pm$ 1.38 
& 2.23 $\pm$ 1.25 
& 2.43 $\pm$ 1.28 
& 2.54 $\pm$ 1.17 
& $[2.11, 2.98]$ 
& $6.155\times 10^{-5}$ & $0.81$ \\

SUA w/o SF
& 100.00
& 16.40 & 14.08 & 51.09 & 82.41
& 6,016 & 5,354 & 11,370
& 5.00 $\pm$ 1.50 
& 4.82 $\pm$ 1.47 
& 4.68 $\pm$ 1.55 
& 4.83 $\pm$ 1.43 
& $[4.30, 5.37]$ 
& $.0667$ & $0.39$ \\

MPA w/o GE
& 100.00
& 18.10 & 11.49 & 46.81 & 77.24
& 6,191 & 5,195 & 11,387
& 5.32 $\pm$ 1.38 
& 4.98 $\pm$ 1.51 
& 4.92 $\pm$ 1.64 
& 5.07 $\pm$ 1.46 
& $[4.53, 5.62]$ 
& $.1304$ & $0.28$ \\

\midrule

\textbf{Full (Ours)}
& 100.00
& 21.09 & 15.94 & 52.86 & 91.38
& 6,439 & 5,457 & 11,896
& \textbf{5.77 $\pm$ 1.15} 
& \textbf{5.47 $\pm$ 1.45} 
& \textbf{5.48 $\pm$ 1.56} 
& \textbf{5.57 $\pm$ 1.35} 
& \textbf{$[5.07, 6.08]$} 
& -- & -- \\

\bottomrule
\end{tabular}
}

\begin{tablenotes}
\small
\item {Note: Ratings range from 1 = Does-Not-Match-At-All to 7 = Fully-Matches and are reported as mean $\pm$ SD over 30 case-level means. 95\% CIs are reported for Avg.\newline \phantom{Note: }Statistical comparisons against Full use two-sided paired Wilcoxon signed-rank tests with Holm correction $p_{\mathrm{Holm}}$; $r$ denotes the corresponding effect size.}
\end{tablenotes}

\end{threeparttable}
\vspace{-2mm}
\end{table*}

\subsubsection{Results}
As shown in Table~\ref{tab:animationeval}, human raters assigned the generated animations an overall score of $5.49 \pm 1.49$. Five-rater judgments showed good average-measure reliability across metrics ($ICC(2,5) > .75$). Motion direction and object-level motion were rated slightly higher than part-level motion and object--part coordination, indicating that fine-grained component execution and coordination remain more challenging than realizing the intended overall behavior.

Category-level results followed the pattern observed during motion planning. Rotating Appliances achieved the highest human rating ($6.18 \pm 0.97$), whereas Aerial Mobility was lowest ($4.84 \pm 1.58$) and Articulated \& Sliding Objects ($5.06 \pm 1.88$) showed the greatest variability. Together with their lower motion-axis accuracy, these results highlight three-dimensional trajectories, asset-specific articulation, and hierarchical dependencies as the primary difficulties in runtime realization. The three LLM judges rated the generated code substantially higher than humans rated the realized animations ($6.46 \pm 0.50$ vs.\ $5.49 \pm 1.49$). This discrepancy suggests that code-level behavioral fidelity does not necessarily translate into equivalent runtime quality, as geometric, articulation, or coordination errors may emerge only during execution.

\subsection{End-to-End Animation Evaluation}
To answer \textbf{TQ1}, we compared the full pipeline with 6 alternatives: (i) \underline{Direct2C} (\textit{Direct-to-Code}) mapped the input directly to Unity code without explicit intermediate representations; (ii) \underline{All-in-One} Agent performed all reasoning and generation within a single agent; (iii) \underline{w/o SUA} (\textit{Semantic Understanding Agent}) removed explicit semantic grounding; (iv) \underline{w/o MPA} (\textit{Motion Planning Agent}) generated code without a structured motion plan; (v) \underline{w/o SF} (\textit{Semantic Frame}) removed explicit guidance for semantic-direction reasoning; and (vi) \underline{w/o GE} (\textit{Geometry Enrichment}) omitted geometry-based parameter estimation from the motion plan. All methods received the same instruction and raw inputs and used the same backbone LLM and decoding settings, with no manual correction.

We report (i) \textbf{Compilation Success}, (ii) \textbf{End-to-End Time}, and (iii) \textbf{Token Consumption} directly from system logs, measuring whether the generated code compiles successfully, the elapsed time from instruction submission to runtime execution, and the input and output tokens consumed by the LLM, respectively. (iv) \textbf{Animation Quality} was assessed through a comparative human evaluation. We selected one prompt for each of the 30 benchmark assets, yielding a 30-case subset balanced across the three prompt levels (10 Simple, 10 Constrained, and 10 Open-Ended), and partitioned the cases into 10 blocks of three. Each block contained one prompt from each level while jointly preserving the benchmark's object-category distribution. 20 independent raters were each assigned one block, such that each block was evaluated by two raters. For each of the three cases, raters evaluated the outputs of all seven methods. Method identities were concealed, and the seven outputs were presented in randomized order. Raters assessed each output on a 7-point scale for \underline{OLM} (\textit{Object-Level Motion}), \underline{PLM} (\textit{Part-Level Motion}), and \underline{OPC} (\textit{Object--Part Coordination}), reflecting the correctness of whole-object behavior, part-level motion, and their coordination.

\subsubsection{Results}
As shown in Table~\ref{tab:execution}, all methods achieved 100\% compilation success, indicating that contemporary LLMs~\cite{openai2026gpt56} can reliably generate syntactically valid Unity C\# code from the provided object and scene context. The main differences therefore lie in efficiency and animation quality rather than compilability. Direct2C was the most efficient, requiring 30.66\,s and 3.93K tokens per case, whereas the full pipeline required 91.38\,s and 11.90K tokens. All-in-One consumed a comparable number of tokens (11.18K) but substantially less time (48.66\,s), suggesting that the additional overhead mainly arises from sequential multi-stage inference.

The full MotionAnimator pipeline achieved the highest overall animation-quality score ($5.57 \pm 1.35$). Holm-corrected paired comparisons showed that our system significantly outperformed Direct2C, All-in-One, w/o SUA, and w/o MPA (all $p_{\mathrm{Holm}} < .01$, $r > .5$). Removing motion planning produced the largest degradation ($2.54 \pm 1.17$, $p_{\mathrm{Holm}} < .001$, $r=.81$), highlighting the importance of explicitly translating semantic understanding into structured motion specifications. Removing the SF reduced the score to $4.83 \pm 1.43$, while removing GE yielded a smaller decrease to $5.07 \pm 1.46$. 
Both variants nevertheless maintained relatively consistent ratings and remained close to the Full pipeline, suggesting that they provide complementary incremental gains within the multi-stage reasoning process. The observed failure patterns further reveal their roles: removing SF led to functionally inconsistent directions, removing GE weakened object--part motion consistency, and removing SUA or MPA increased incorrect motion axes, directions, and missing part motions. Direct2C nevertheless occasionally produced plausible behaviors, such as subtle hovering motions for aerial objects, suggesting that LLMs retain useful motion priors even without explicit intermediate representations, but apply them less reliably across objects and instructions. Overall, explicit semantic and motion reasoning---particularly structured motion planning---improved the consistency and alignment of generated animations relative to the baseline and ablated variants.

\begin{figure}[t]
    \centering
    \includegraphics[width=1.0\linewidth]{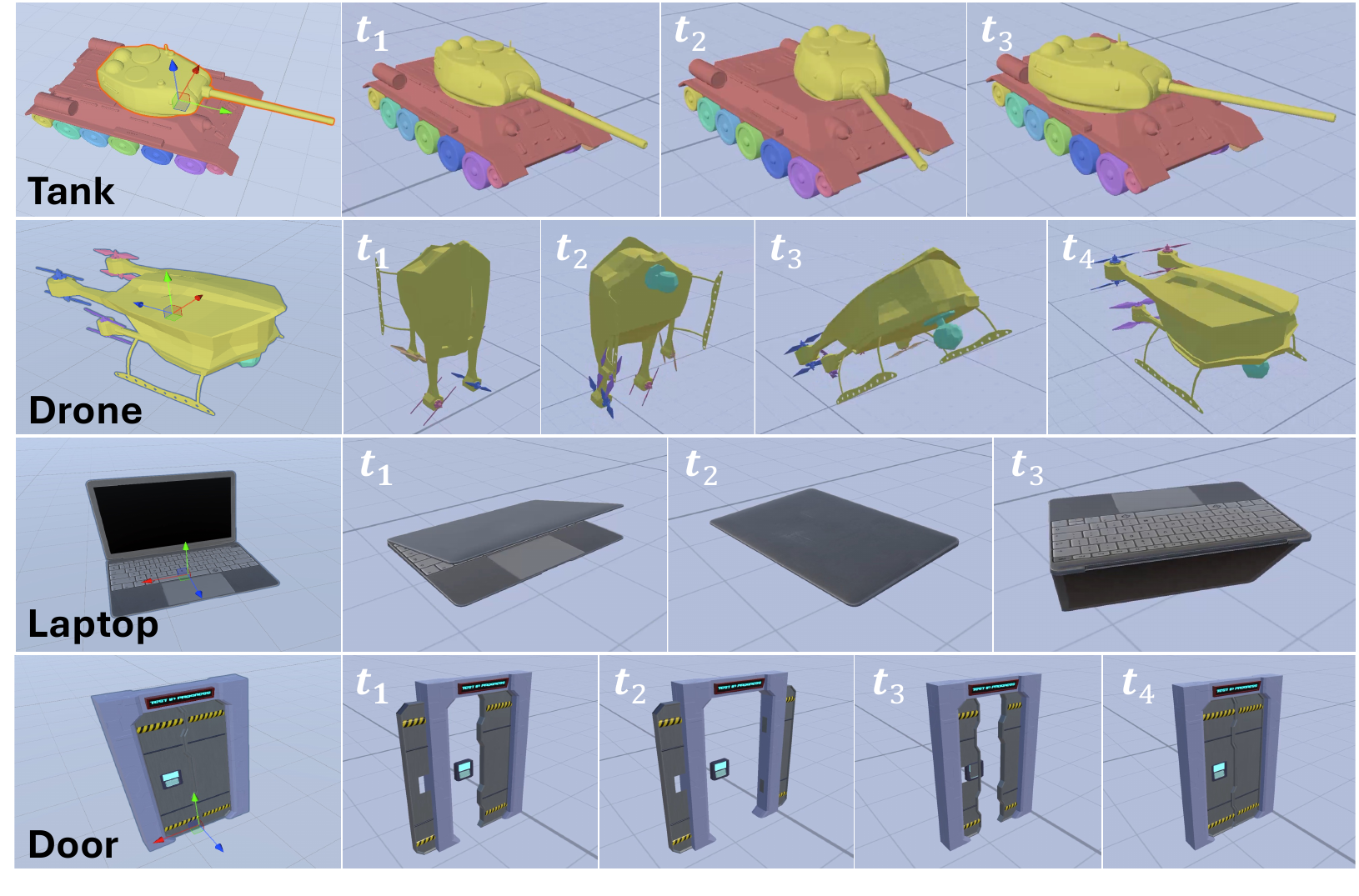}
    \vspace{-6mm}
    \caption{Representative failure cases of generated animations.}
    \label{fig:failure}
    \vspace{-2mm}
\end{figure}

\subsection{Failure Analysis}
As shown in Fig.~\ref{fig:failure}, we observed four representative failure modes. (i) For irregularly shaped movable parts, the system may estimate an inaccurate motion center when the asset does not provide a suitable geometric pivot, resulting in incorrect rotations such as the tank turret case. (ii) Motions involving takeoff or landing can be underspecified by high-level language; without explicit trajectory constraints, the system may generate implausible ascent or descent behaviors, as shown for the drone. (iii) Although the system can identify a plausible rotation axis, it may fail to infer an appropriate rotation range, causing excessive articulation, such as the laptop screen rotating beyond its physically plausible range. (iv) Part-motion dependencies are not always fully captured. A component attached to a moving part may remain stationary if the system does not infer that it should follow the parent motion, as illustrated by the control panel on the sliding door. These cases highlight remaining challenges in geometric pivot estimation, motion-range reasoning, trajectory specification, and hierarchical part coordination.



\section{User Study}
The technical evaluation assessed how well MotionAnimator translates animation intent into coordinated object--part behavior. We further conducted an exploratory user study to investigate how people interact with this semantic-to-motion workflow.

\subsection{Study Goals}
The study was guided by three questions:

\vspace{-2mm}
\begin{itemize}[leftmargin=9mm, itemsep=-1mm]
    \item[\textbf{UQ1}:] \textbf{Intent Formation and Evolution.}
    \textit{At what level of abstraction do users formulate animation intent, and how does it evolve through interaction with generated behavior?}

    \item[\textbf{UQ2}:] \textbf{Interpretation and Refinement.}
    \textit{How do users use part semantics and motion plans to interpret generated animations, diagnose mismatches, and formulate revisions?}

    \item[\textbf{UQ3}:] \textbf{Opportunities and Limitations.}
    \textit{How do students, instructors, and XR developers perceive the potential uses and limitations of intent-driven, part-aware animation authoring?}
\end{itemize}
\vspace{-2mm}

\subsection{Participants}
We recruited 9 participants (3 female, 6 male; age range 20--28, $M=24.0$) from three user groups: 3 students, 3 instructors, and 3 XR developers. Participants had diverse academic backgrounds spanning science, mathematics, engineering, and computing disciplines. All reported prior XR experience. Students represented potential users of animated 3D content for learning; instructors, creators of dynamic 3D demonstrations for teaching; and XR developers, users of generated object behaviors for rapid prototyping. The study was approved by our university IRB, and each participant received a \$15 Amazon gift card as compensation.

\subsection{Procedure}
\begin{figure}[t]
    \centering
    \includegraphics[width=1.0\linewidth]{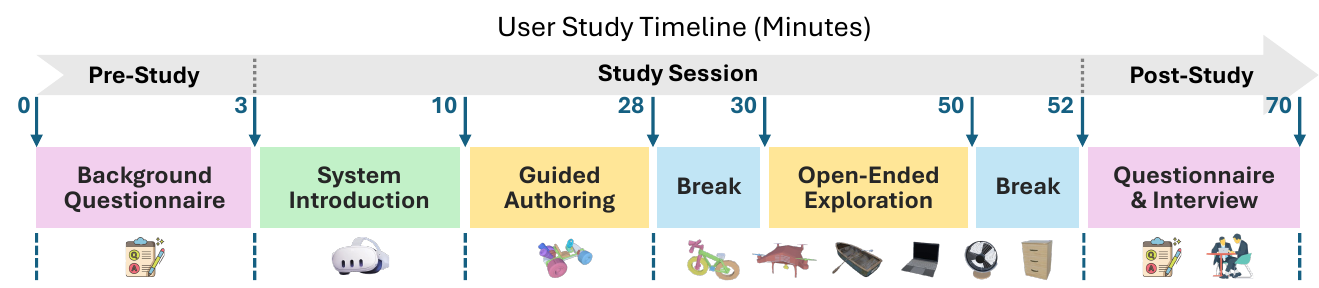}
    \vspace{-6mm}
    \caption{User Study Timeline.}
    \vspace{-2mm}
    \label{fig:timeline}
\end{figure}

Each study session lasted approximately 70 minutes and consisted of pre-study preparation, guided interaction, open-ended exploration, and post-study feedback, as illustrated in Figure~\ref{fig:timeline}.

\begin{table}[t]
\centering
\caption{Post-study questionnaire and semi-structured interview.}
\label{tab:questionnaire}
\fontsize{6pt}{6pt}\selectfont
\setlength{\tabcolsep}{3pt}
\renewcommand{\arraystretch}{1.08}
\begin{tabularx}{\linewidth}{@{}c@{\hspace{1mm}}X@{}}
\toprule
\textbf{Item} & \textbf{Statement / Guiding Question} \\
\midrule

\multicolumn{2}{l}{\textbf{Post-study Questionnaire} (7-point Likert scale (1 = Strongly Disagree, 7 = Strongly Agree))} \\

\multicolumn{2}{l}{\subarrow \textit{Usability and Expression}} \\

\textbf{Q1} &
The system was easy to understand and use. \\

\textbf{Q2} &
I could express my desired animation using natural language. \\

\textbf{Q3} &
I could describe animations in terms of object function rather than low-level motion parameters. \\

\midrule
\multicolumn{2}{l}{\subarrow \textit{Perceived Animation Quality}} \\

\textbf{Q4} &
The generated animations matched my intended whole-object motion. \\

\textbf{Q5} &
The generated animations matched my expected part-level motion. \\

\textbf{Q6} &
The generated animations coordinated whole-object and part-level motion in a plausible way. \\

\midrule
\multicolumn{2}{l}{\subarrow \textit{Understanding and Refinement}} \\

\textbf{Q7} &
The semantic part labels helped me understand how the system interpreted the object. \\

\textbf{Q8} &
The motion plan helped me understand how my instruction was translated into animation. \\

\textbf{Q9} &
When an animation was imperfect, the semantic part labels or motion plan helped me identify what went wrong. \\

\textbf{Q10} &
I could refine the animation when the initial result did not match my intent. \\

\textbf{Q11} &
I felt I had control over which parts moved and how they moved. \\

\midrule
\multicolumn{2}{l}{\subarrow \textit{Potential Applications}} \\

\textbf{Q12} &
The system would be useful for creating educational or explanatory XR demonstrations. \\

\textbf{Q13} &
The system would be useful for early-stage XR prototyping. \\

\midrule

\multicolumn{2}{l}{\textbf{Post-study Semi-structured Interview}} \\

\multicolumn{2}{l}{\subarrow \textit{Authoring Experience}} \\

\textbf{I1} &
How did you decide what prompt to give the system? Did you think in terms of object function, whole-object motion, individual parts, or low-level motion parameters? \\

\textbf{I2} &
How did it feel to give a broad functional instruction compared with a more specific coordinated motion instruction? \\

\textbf{I3} &
In the free exploration task, why did you choose that object and animation? \\

\midrule
\multicolumn{2}{l}{\subarrow \textit{Interpretation and Refinement}} \\

\textbf{I4} &
What made a generated animation feel correct or incorrect to you? Did you pay more attention to whole-object motion, part-level motion, or their coordination? \\

\textbf{I5} &
Did the semantic part labels help you understand how the system interpreted the object? Why or why not? \\

\textbf{I6} &
Did the motion plan help you understand or refine the generated animation? Why or why not? \\

\textbf{I7} &
When the result did not match your intent, how did you decide what to revise? \\

\midrule
\multicolumn{2}{l}{\subarrow \textit{Applications and Limitations}} \\

\textbf{I8} &
Where would this system be useful today, and what would need to improve before you would rely on it in a real class, project, prototype, or demonstration? \\

\bottomrule
\end{tabularx}
\vspace{-4mm}
\end{table}

Participants first completed a short \myul{Background Questionnaire} about their prior experience with XR, 3D modeling, AI-authoring, and Unity or game development. We also asked whether they had previously created, explained, or prototyped dynamic 3D content in their learning, teaching, or development activities.

Following the questionnaire, the experimenter introduced the basic MotionAnimator \myul{Workflow}: selecting a part-structured 3D object, providing a natural-language animation instruction, previewing the generated animation, inspecting the semantic part labels and motion plan, and revising the animation when desired. The introduction focused on how to operate the system without prescribing how participants should formulate their instructions.

Participants then completed two \myul{Guided Interactions} using a \textit{Car}. First, they provided the instruction ``\textit{Move forward}'' to become familiar with the workflow. They were then shown a pre-generated ``\textit{Move in an S-shape}'' animation containing a deliberate part-motion error: the car followed the requested trajectory, but the steering wheel rotated around an incorrect axis. Participants were asked to identify the mismatch, inspect the semantic part labels and motion plan, and formulate a revision. The second task provided a common diagnostic case for examining how participants interpreted and refined an imperfect animation.

Participants then proceeded to an \myul{Open-Ended Exploration} using six representative objects---\textit{Bicycle}, \textit{Drone}, \textit{Rowboat}, \textit{Table Fan}, \textit{Laptop}, and \textit{Drawer}---covering ground and aerial locomotion, coordinated articulation, continuous rotation, hinge articulation, and linear sliding. Participants could freely choose objects, create animations using their own instructions, inspect the intermediate representations, and revise the behaviors as desired. No target animations were prescribed, allowing participants to decide both what behaviors to create and how specifically to describe them.

Finally, participants completed an \myul{Experience Questionnaire} and participated in a \myul{Semi-Structured Interview} about their authoring experience, interpretation and refinement of generated animations, and potential applications and limitations of the system.

\subsection{Measures and Data Analysis}
Participants completed 13 questionnaire items assessing their overall experience with MotionAnimator and subsequently participated in a semi-structured interview to further understand their authoring, interpretation, and refinement experiences. Table~\ref{tab:questionnaire} summarizes the questionnaire items and interview questions. Questionnaire responses were analyzed descriptively using means and standard deviations. We conducted a descriptive analysis of participants' interview responses to identify recurring observations, which contextualize the questionnaire results and organize the qualitative findings.

\begin{figure}[t]
    \centering
    \includegraphics[width=1.0\linewidth]{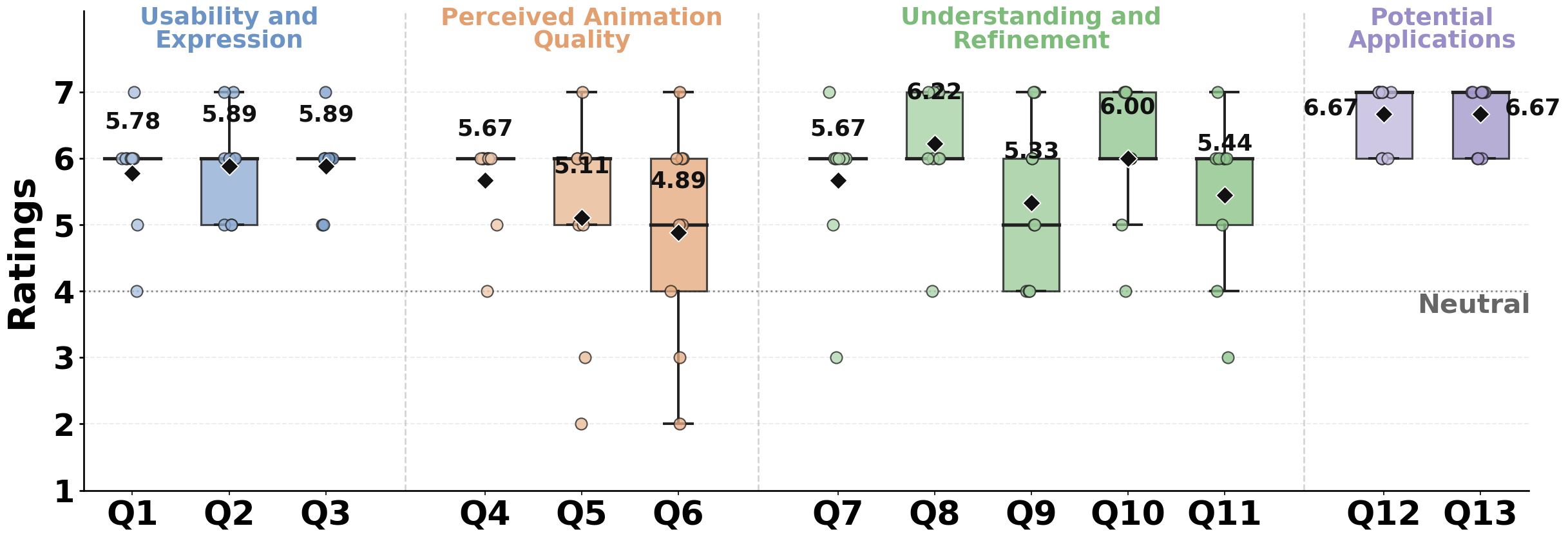}
    \vspace{-6.5mm}
    \caption{Post-Study questionnaire results.}
    \vspace{-5.5mm}
    \label{fig:questionnaire}
\end{figure}

\subsection{Results and Findings}
Fig.~\ref{fig:questionnaire} shows overall positive post-study ratings across four aspects.

\textbf{F1: Animation authoring emerged as a process of progressive specification across levels of abstraction.}
Participants found the system easy to understand and use ($M=5.78 \pm 0.83$; Q1), and reported that they could express desired animations using natural language ($M=5.89 \pm 0.78$; Q2) and describe behavior through object function rather than low-level motion parameters ($M=5.89 \pm 0.60$; Q3). Interviews showed that participants did not rely on a fixed level of specification, but moved toward more detailed constraints as their intended behavior became clearer. P4 described starting with ``\textit{whole-object motion at first rather than individual parts},'' whereas P8 noted that sometimes they had ``\textit{only broad thoughts},'' making broad instructions a useful starting point. P9 similarly explained, ``\textit{I initially focused on whole-object motion, but during later refinement I adjusted the animation according to the mobility of each part}.'' Consistent with this progression, whole-object motion was rated positively ($M=5.67 \pm 0.71$; Q4), while part-level motion ($M=5.11 \pm 1.62$; Q5) and object--part coordination ($M=4.89 \pm 1.62$; Q6) showed greater variability. These results suggest that high-level intent provides an accessible starting point, while progressively introducing part- and coordination-level constraints supports finer-grained specification as authoring proceeds.

\textbf{F2: Participants used intermediate representations as complementary references for interpretation, diagnosis, and refinement.}
Participants found semantic part labels useful for understanding how the system interpreted an object ($M=5.67 \pm 1.12$; Q7), while the motion plan received particularly strong ratings for explaining how an instruction was translated into animation ($M=6.22 \pm 0.97$; Q8). Participants also reported positive experiences in identifying problems ($M=5.33 \pm 1.22$; Q9), refining imperfect results ($M=6.00 \pm 1.00$; Q10), and controlling which parts moved and how ($M=5.44 \pm 1.24$; Q11). P1 noted that the motion plan helped ``\textit{understand how each step was generated},'' while P6 found its summary most useful for ``\textit{understanding how the system reasons}.'' Complementarily, P7 reported that the intermediate information provided ``\textit{a clear idea of each component}'' and its animation path when deciding what to revise. These accounts suggest that participants used semantic labels as component-level references and motion plans as behavioral references when inspecting and refining generated animations.

\textbf{F3: Participants viewed the system as a medium for rapid behavioral exploration as well as animation generation.}
Potential applications received the highest ratings for educational or explanatory XR demonstrations ($M=6.67 \pm 0.50$; Q12) and early-stage XR prototyping ($M=6.67 \pm 0.50$; Q13). Participants envisioned applications spanning both learning and design. P2 described its potential for helping learners understand ``\textit{how real objects work},'' while P5 highlighted ``\textit{object architecture and prototype design}'' before physical development. During free exploration, participants also used the system to probe behaviors beyond straightforward reproduction: P3 deliberately selected a rowboat because it offered ``\textit{more degrees of freedom},'' allowing exploration of how its parts moved relative to the whole object. Other explorations included simulating a fan moving backward under airflow reaction forces, exploding a bicycle to inspect its component structure, and creating a rocking motion for a boat on the water. These observations suggest that participants valued the system not only for producing a predetermined animation, but also for quickly testing relationships among object structure, part behavior, and motion.

\subsection{Discussion and Limitations}
Three limitations currently define the system's animation scope.
\textbf{First}, geometric grounding limits motion flexibility. MotionAnimator currently relies largely on existing local axes and geometric centers when determining motion axes and pivots. Thus, a semantically plausible plan may still produce incorrect articulation when these geometric assumptions do not match the intended motion. Future systems could infer instruction-conditioned articulation frames, including part-specific pivots, axes, and motion ranges, from geometry, hierarchy, semantics, and user intent. 
\textbf{Second}, iterative refinement can preserve unwanted motion assumptions. Revisions reuse prior motion context for efficient incremental modification, but plausible yet unintended behavior may persist and accumulate across revisions. Future revision agents could explicitly determine which prior constraints to preserve, replace, or remove, and selectively re-plan affected motion dependencies. 
\textbf{Third}, the current animation space remains object-centric. The system primarily animates existing objects and their constituent parts, limiting behaviors that require effects beyond object motion itself. Participants' explorations also pointed toward richer physical and environmental phenomena. Future systems could extend part-aware motion generation beyond existing object geometry to incorporate generated effects and dynamic content, such as exhaust, airflow, or animated displays, enabling richer and more complete animations.

\section{Conclusion}
In this work, we introduce MotionAnimator, which reframes XR object animation as semantic motion grounding: resolving high-level user intent into explicit asset semantics and coordinated object- and part-level motion before runtime realization. Our results show that executability alone is insufficient, as methods producing compilable Unity animations differed substantially in animation correctness and coordination. By externalizing semantic and motion decisions into asset-grounded intermediate representations, MotionAnimator enables otherwise inert part-structured assets to support coordinated behaviors while making the reasoning from user intent to runtime execution traceable. More broadly, our work suggests that future generative XR systems should not only produce content, but also expose the intermediate structures through which user intent becomes situated, inspectable, and executable behavior.

\bibliographystyle{abbrv-doi}

\bibliography{main}
\end{document}